\documentclass[12pt]{article}
\usepackage[latin1]{inputenc}
\usepackage[dvips]{graphicx}
\usepackage{graphicx}
\newcommand{\minrm}{{\rm min}}
\newcommand{\maxrm}{{\rm max}}

\newcommand{\drm}{{\rm d}}

\newcommand{\text}{\rm}

\newcommand{\ug}{ \; = \; }

\newcommand{\la}{\lambda}
\newcommand{\La}{\Lambda}

\newcommand{\bb}{\begin{equation}}
\newcommand{\ee}{\end{equation}}
\newcommand{\bega}{\begin{eqnarray}}
\newcommand{\ega}{\end{eqnarray}}
\newcommand{\begae}{\begin{eqnarray*}}
\newcommand{\egae}{\end{eqnarray*}}

\newcommand{\h}{\hspace*{4ex}}
\newcommand{\dis}{\displaystyle}

\newcommand{\be}{\beta}

\newcommand{\om}{\omega}

\newcommand{\cent}{\centerline}
\newcommand{\vs}{\vspace*}

\begin{document}

\baselineskip 0.8cm

\begin{center}

{\large {\bf Theory of Frozen Waves}$^{\: (\dag)}$}
\footnotetext{$^{\: (\dag)}$  Work partially supported by MIUR and INFN
(Italy), and by FAPESP (Brazil). \ This paper did first appear as e-print
physics/*******. \ E-mail addresses for contacts:
mzamboni@dmo.fee.unicamp.br; \ recami@mi.infn.it}

\end{center}

\vs{5mm}

\cent{ M. Zamboni-Rached, }

\vs{0.2 cm}

\centerline{{\em DMO--FEEC, State University at Campinas,
Campinas, SP, Brazil.}}

\vs{0.5 cm}

\cent{ Erasmo Recami }

\vs{0.2 cm}

\cent{{\em Facolt\`a di Ingegneria, Universit\`a statale di Bergamo,
Dalmine (BG), Italy;}}
\cent{{\rm and} {\em INFN---Sezione di Milano, Milan, Italy.}}

\vs{0.2 cm}

\centerline{\rm and}

\vs{0.3 cm}

\cent{ H. E. Hern\'{a}ndez-Figueroa }

\vs{0.2 cm}

\cent{{\em DMO--FEEC, State University at Campinas,
Campinas, SP, Brazil.}}

\vs{0.5 cm}

{\bf Abstract  \ --} \ In this work, starting by suitable superpositions
of equal-frequency Bessel beams, we develop a theoretical and
experimental methodology to obtain localized {\em stationary} wave
fields, with high transverse localization, {\em whose longitudinal
intensity pattern can approximately assume any desired shape} within a chosen
interval $0\leq z \leq L$ of the propagation axis $z$. Their intensity
envelope remains static, i.e. with velocity $v=0$; so that we have named
``Frozen Waves" (FW) these new solutions to the wave equations (and, in
particular, to the Maxwell equations). Inside the envelope of a FW only
the carrier wave does propagate: And the longitudinal shape,
within the interval $0\leq z \leq L$, can be chosen in such a way that
no nonnegligible field exists outside the pre-determined region (consisting,
e.g., in one or more high intensity peaks). Our solutions are noticeable
also for the different and interesting applications they can have, especially
in electromagnetism and acoustics, such as
optical tweezers, atom guides, optical or acoustic bistouries, various
important medical apparata, etc.

{\em PACS nos.}: \ 41.20.Jb ; \ 03.50.De ; \ 03.30.+p ; \
84.40.Az ; \ 42.82.Et ; \ 83.50.Vr ; \ \ 62.30.+d ; \
43.60.+d ; \  91.30.Fn ; \  04.30.Nk ; \  42.25.Bs ; \ 46.40.Cd ; \
52.35.Lv \ .\hfill\break



\newpage

\

\section{Introduction}

Over many years the theory of localized waves (LW), or
nondiffracting waves, and in particular of the so-called X-shaped
waves, has been developed, generalized,
and experimentally verified in many fields, such as optics,
microwaves and acoustics\cite{1}. These new solutions
to the wave equations (and, in particular, to the Maxwell equations)
have the noticeable
characteristic of resisting the diffraction effects for long
distances, i.e., of possessing a large depth of field.

\h Such waves can be divided into two classes: the localized beams
and the localized pulses.  With regard to the beams, the most
popular is the Bessel beam.

\h Much work was made about the properties and applications of
a single Bessel beam, while some work has been done in connection with
Bessel beam superpositions performed by summing or integrating over their
frequency (by producing, e.g., the well-known ``X-shaped pulses"
and/or their velocity (for example,
it has been studied the space-time focusing of different-speed X-shaped
pulses).
By contrast, only a few papers have been
addressed to the properties and applications of {\em
superpositions} of Bessel beams {\em with the same frequency, but} with
different longitudinal wave numbers. The few works existing on this subject
have shown some surprising possibilities associated with this
particular type of
superpositions, mainly the possibility of controlling the
transverse shape of the resulting beam\cite{b1,b2}.

\h The other
important point, i.e., that of controlling the longitudinal shape,
has been even more rarely addressed, and the relevant papers have been
so far confined to numerical optimization processes\cite{ol,on}, aimed at
finding out an appropriate computer-generated hologram.

\h In this work we develop a very simple method$^{**}$
\footnotetext{$^{**}$ Patent pending.}, having recourse to superpositions
of forward propagating and equal-frequency Bessel beams only, that allows
controlling the beam-intensity {\em longitudinal} shape within a chosen
interval $0\leq z \leq L$, where $z$ is the propagation axis and
$L$ can be much greater than the wavelength $\lambda$ of the
monochromatic light (or sound) which is being used. Inside such a space
interval, indeed, we succeed in constructing a {\em stationary} envelope whose
longitudinal intensity pattern can approximately assume any desired shape,
including, for instance, one or more high-intensity peaks (with
distances between them much larger than $\lambda$); and which results ---in
addition--- to be naturally endowed also with a good transverse
localization.$^{***}$
\footnotetext{$^{***}$ When we get a complete control on the longitudinal
shape, we cannot have ---however--- a total control also on the transverse
localization, since our stationary fields are of course constrained to obey
the wave equation.}

\h This intensity
envelope remains static, i.e., has velocity $v=0$; and because of this
in a previous paper\cite{Mic} we have called ``Frozen Waves" (FW)
these new solutions to the wave equations (and, in
particular, to the Maxwell equations). Inside the envelope of a FW only
the carrier wave does propagate: And the longitudinal shape,
within the interval $0\leq z \leq L$, can be chosen in such a way that
no nonnegligible field exists outside the pre-determined high-intensity
region.

\h We also suggest a simple apparatus capable of generating the mentioned
stationary fields.

\h Static wave solutions like these are noticeable also for the
different and interesting applications they can have, especially
in electromagnetism and acoustics, such as optical tweezers, atom
guides, optical or acoustic bistouries, optical micro-lithography,
electromagnetic or ultrasound high-intensity fields for various
important medical purposes, and so on.$^{**}$

\

\

\section{The mathematical methodology:$^{**}$
Stationary wavefields with arbitrary longitudinal shape,
obtained by superposing equal-frequency Bessel beams}

Let us start from the well-known axis-symmetric zeroth order Bessel beam
solution to the wave equation:

\bb \psi(\rho,z,t)\ug J_0(k_{\rho}\rho)e^{i\be z}e^{-i\om t}
\label{bb}\ee

with

\bb k_{\rho}^2=\frac{\om^2}{c^2} - \be^2 \; , \label{k}  \ee

where
$\om$, $k_{\rho}$ and $\be$ are the angular frequency, the
transverse and the longitudinal wave numbers, respectively. We
also impose the conditions

\bb \om/\be > 0 \;\;\; {\rm and}\;\;\; k_{\rho}^2\geq 0
\label{c2} \ee

(which imply $\om/\be \geq c$) to ensure forward propagation only (with
no evanescent waves), as well
as a physical behavior of the Bessel function $J_0$.

\h Now, let us make a superposition of $2N + 1$ Bessel beams with the
same frequency $\om_0$, but with {\em different} (and still
unknown) longitudinal wave numbers $\be_n$:

\bb \dis{\Psi(\rho,z,t) \ug e^{-i\,\om_0\,t}\,\sum_{n=-N}^{N}
A_n\,J_0(k_{\rho\,n}\rho)\,e^{i\,\be_n\,z} } \; , \label{soma} \ee

where $n$ are integer numbers and $A_n$ are constant coefficients.
For each $n$, the
parameters $\om_0$, $k_{\rho\,n}$ and $\beta_n$ must satisfy
Eq.(\ref{k}), and, because of conditions (\ref{c2}), when considering
$\om_0 > 0$, we must have

\bb 0 \leq \be_n \leq \frac{\om_0}{c} \; . \label{be} \ee

\h Let us now suppose that we wish $|\Psi(\rho,z,t)|^2$ of
Eq.(\ref{soma}) to assume on the axis $\rho=0$ the pattern
represented by a function $|F(z)|^2$, inside the chosen interval
$0 \leq z \leq L$. In this case, the function $F(z)$ can be
expanded, as usual, in a Fourier\footnote{Such a choice of the
longitudinal intensity pattern does imply an interesting freedom,
since we can consider more in general any expansion
$\sum_{m=-\infty}^{\infty}\,B_m\,\exp{{i\,\frac{2\pi}{L}\,m\,z}} =
F(z)\,\exp{{i\phi (z)}}$, quantity $\phi (z)$ being an arbitrary function of
the coordinate $z$.} series:

\

$$F(z) \ug
\sum_{m=-\infty}^{\infty}\,B_m\,e^{i\,\frac{2\pi}{L}\,m\,z} \; ,$$

\

where

\

$$B_m \ug \frac{1}{L} \dis{
\int_{0}^{L}\,F(z)\,e^{-i\,\frac{2\pi}{L}\,m\,z}\,\drm\,z } \ .$$

\

\noindent More precisely, our goal is finding out, now, the values
of the longitudinal wave numbers $\be_n$ and of the coefficients
$A_n$, of Eq.(\ref{soma}), in order to reproduce approximately,
within the said interval $0 \leq z \leq L$ (for $\rho=0$), the
predetermined longitudinal intensity-pattern $|F(z)|^2$. \ Namely,
we want to have

\bb \left|\,\sum_{n=-N}^{N} A_n e^{i\,\be_n\,z}\right|^{\,2}
\approx |F(z)|^{\,2} \;\;\;\; {\rm with}\;\;\; 0\leq z \leq L  \;
. \label{soma1} \ee

\

\h Looking at Eq.(\ref{soma1}), one might be tempted to take
$\be_n = 2\pi n/L$, thus obtaining a truncated Fourier series,
expected to represent approximately the desired pattern $F(z)$. \
Superpositions of Bessel beams with $\be_n = 2\pi n/L$ have been
actually used in some works to obtain a large set of {\it
transverse} amplitude profiles\cite{b1,b2}. However,
for our purposes, this choice is not appropriate, due to two
principal reasons: \ 1) It yields negative values for $\be_n$
(when $n<0$), which implies backwards propagating components
(since $\om_0 > 0$); \ 2) In the cases when $L>>\lambda_0$, which
are of our interest here, the main terms of the series would
correspond to very small values of $\be_n$, which results in a
very short field-depth of the corresponding Bessel beams (when
generated by finite apertures), preventing the creation of the
desired envelopes far form the source.

\h Therefore, we need to make a better choice for the values of
$\be_n$, which allows forward propagation components only, and a
good depth of field. \ This problem can be solved by putting

\bb \be_n \ug Q + \frac{2\,\pi}{L}\,n \; , \label{be2}  \ee

where $Q>0$ is a value to be chosen (as we shall see) according to the
given experimental situation, and the desired degree of {\em transverse}
field localization. \ Due to Eq.(\ref{be}), we get

\bb 0\leq Q \pm \frac{2\,\pi}{L}\,N \leq \frac{\om_0}{c} \; . \label{N}
\ee

\h Inequality (\ref{N}), can be used to determine the maximum
value of $n$, that we call $N_{\rm max}$, once $Q$, $L$ and
$\om_0$ have been chosen.

\h As a consequence, for getting a longitudinal intensity pattern
approximately equal to the desired one, $|F(z)|^2$, in the
interval $0\leq z \leq L $, Eq.(\ref{soma}) should be rewritten as

\bb \dis{\Psi(\rho=0,z,t) \ug
e^{-i\,\om_0\,t}\,e^{i\,Q\,z}\,\sum_{n=-N}^{N}
A_n\,e^{i\,\frac{2\pi}{L}n\,z} } \; , \label{soma2} \ee

with

\bb A_n \ug \frac{1}{L} \dis{
\int_{0}^{L}\,F(z)\,e^{-i\,\frac{2\pi}{L}\,n\,z}\,\drm\,z } \; .
\label{An} \ee

Obviously, one obtains only an approximation to the desired
longitudinal pattern, because the trigonometric series
(\ref{soma2}) is necessarily truncated ($N \leq N_{\rm max}$). Its
total number of terms, let us repeat, will be fixed once the
values of $Q$, $L$ and $\om_0$ are chosen.

\h When $\rho \neq 0$, the wavefield $\Psi(\rho,z,t)$ becomes

\bb \dis{\Psi(\rho,z,t) \ug
e^{-i\,\om_0\,t}\,e^{i\,Q\,z}\,\sum_{n=-N}^{N}
A_n\,J_0(k_{\rho\,n}\,\rho)\,e^{i\,\frac{2\pi}{L}n\,z} } \; ,
\label{soma3} \ee

with

\bb k_{\rho\,n}^2 \ug \om_0^2 - \left(Q + \frac{2\pi\,n}{L}
\right)^2 \; . \label{krn} \ee

\h The coefficients $A_n$ will yield {\em the amplitudes} and {\em the
relative phases} of each Bessel beam in the superposition.

\h Because we are adding together zero-order Bessel functions, we
can expect a {\em high} field concentration around $\rho=0$.
Moreover, due to the known non-diffractive behavior of the Bessel
beams, we expect that the resulting wavefield will preserve its
transverse pattern in the entire interval $0\leq z \leq L $.

\h The methodology developed here deals with the longitudinal
intensity pattern control. Obviously, we cannot get a total 3D
control, due the fact that the field must obey the wave equation.
However, we can use two ways to have some control over the
transverse behavior too. The first is through the parameter $Q$ of
Eq.(7). \ Actually, we have some freedom in the choice of
this parameter, and FWs representing the same longitudinal
intensity pattern can possess different values of $Q$. The
important point is that, in superposition (\ref{soma3}), using
a smaller value of $Q$ makes the Bessel beams possess a higher
transverse concentration (because, on decreasing the value of $Q$, one
increases the value of the Bessel beams transverse numbers),
and this will reflect in the resulting field, which will present
a narrower central transverse spot. We will exemplify this
fact in the next Section. \ The second way to control the
transverse intensity pattern is using higher order Bessel beams,
but we shall show this in Section 5.

\

\

\section{Some examples}

In this Section we shall present a few examples of our methodology.

\

\h {\em First example:}

Let us suppose that we want an optical wavefield with $\lambda_0
= 0.632\;\mu$m, that is, with $\om_0 = 2.98 \times 10^{15}\;$Hz,
whose longitudinal pattern (along its $z$-axis) in the range $0
\leq z \leq L$ is given by the function

 \bb
 F(z) \ug \left\{\begin{array}{clr}
 -4\,\,\dis{\frac{(z-l_1)(z-l_2)}{(l_2 - l_1)^2}} \;\;\; & {\rm for}\;\;\; l_1 \leq z \leq l_2  \\
\\
 \;\;\;\;\;\;\;\;1 & {\rm for}\;\;\; l_3 \leq z \leq l_4 \\
\\
 -4\,\,\dis{\frac{(z-l_5)(z-l_6)}{(l_6 - l_5)^2}} & {\rm for}\;\;\; l_5 \leq z \leq
 l_6 \\
 \\
 \;\;\;\;\;\;\;\; 0  & \mbox{elsewhere} \ ,
\end{array} \right. \label{Fz1}
 \ee


where $l_1=L/5-\Delta z_{12}$ and $l_2=L/5+\Delta z_{12}$ with
$\Delta z_{12}=L/50$; while $l_3=L/2-\Delta z_{34}$ and $l_4=L/2+\Delta
z_{34}$ with $\Delta z_{34}=L/10$; and, at last, $l_5=4L/5-\Delta z_{56}$
and $l_6=4L/5+\Delta z_{56}$ with $\Delta z_{56}=L/50$. In other
words, the desired longitudinal shape, in the range $0 \leq z \leq
L$, is a parabolic function for $l_1 \leq z \leq l_2$, a unitary
step function for $l_3 \leq z \leq l_4$, and again a parabola in
the interval $l_5 \leq z \leq l_6$, it being zero elsewhere (within
the interval $0 \leq z \leq L$, as we said). In this example, let
us put $L=0.2\;$m.

\h We can then easily calculate the coefficients $A_n$, which
appear in the superposition (\ref{soma3}), by inserting
Eq.(\ref{Fz1}) into Eq.(\ref{An}). Let us choose, for instance,
$Q=0.999\,\om_0/c$: This choice allows the maximum value
$N_{\rm max}=316$ for $n$, as one can infer from Eq.(\ref{N}).  Let us
emphasize that one is not compelled to use just
$N=316$, but can adopt for $N$ any values {\em smaller} than
it; more generally, any value smaller than that calculated via
Eq.(\ref{N}). Of course, on using the maximum value allowed for
$N$, one gets a better result.

\h In the present case, let us adopt the value $N=30$. In Fig.1(a)
we compare the intensity of the desired longitudinal function
$F(z)$ with that of the Frozen Wave, \ $\Psi(\rho=0,z,t)$, \
obtained from Eq.(\ref{soma2}) by adopting the mentioned value
$N=30$.

\

\begin{figure}[!h]
\begin{center}
 \scalebox{1}{\includegraphics{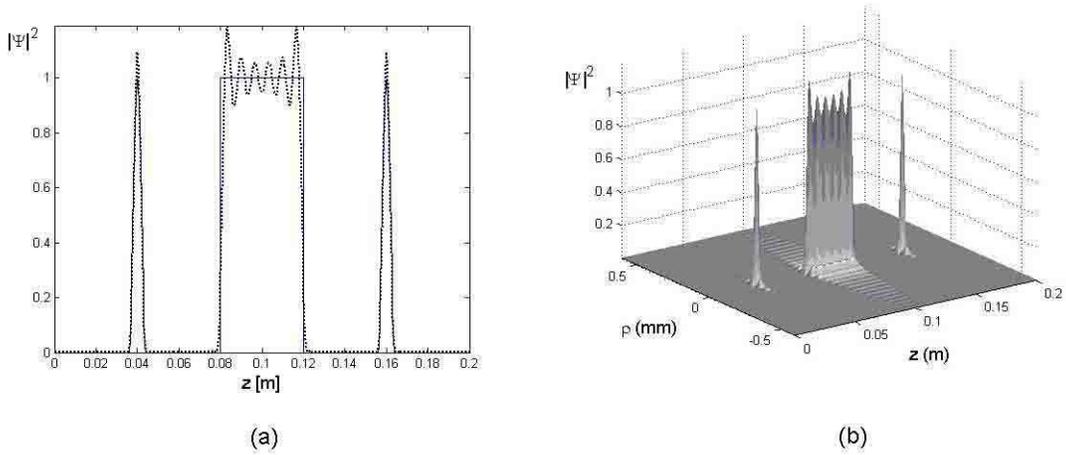}}
\end{center}
\caption{\textbf{(a)} Comparison between the intensity of the
desired longitudinal function $F(z)$ and that of our Frozen Wave
(FW), \ $\Psi(\rho=0,z,t)$, \ obtained from Eq.(\ref{soma2}). The
solid line represents the function $F(z)$, and the dotted one our
FW. \textbf{(b)} 3D-plot of the field-intensity of the FW chosen
in this case by us.} \label{fig1}
\end{figure}

One can verify that a good agreement between the desired
longitudinal behavior and our approximate FW is already got
with $N=30$. The use of higher values for $N$ can only improve the
approximation. \ Figure 1(b) shows the 3D-intensity of our FW, given by
Eq.(\ref{soma3}). One can observe that this field possesses the
desired longitudinal pattern, while being endowed with a good
transverse localization.

\

\h {\em Second example:}

Let us now suppose we want an optical wavefield with
$\la=0.632\;\mu$m ($\om_0=2.98\,10{15}\;$Hz), whose longitudinal
pattern (on its axis) in the range $0 \leq z \leq L$ consists in a
pair of parabolas, for $l_1 \leq z \leq l_2$ and $l_3 \leq z \leq
l_4$, the intensity of the second parabola being twice as much as
that of the first one. Outside the intervals $l_1 \leq z \leq
l_2 \;\; \bigcup \;\; l_3 \leq z \leq l_4$, the desired field has
a null intensity. Summarizing, we want :

\bb
 F(z) \ug \left\{\begin{array}{clr}
 -4\,\,\dis{\frac{(z-l_1)(z-l_2)}{(l_2 - l_1)^2}} \;\;\; & {\rm for}\;\;\; l_1 \leq z \leq l_2  \\
\\
 -4\sqrt{2}\,\,\dis{\frac{(z-l_3)(z-l_4)}{(l_4 - l_3)^2}} & {\rm for}\;\;\; l_3 \leq z \leq
 l_4 \\
 \\
 \;\;\;\;\;\;\;\; 0  & \mbox{elsewhere} \ ,
\end{array} \right. \label{par}
 \ee

where $l_1=3L/10 -\Delta z_{12}$ and $l_2=3L/10+\Delta z_{12}$ with
$\Delta z_{12}=L/70$; while $l_3=7L/10-\Delta z_{34}$ and $l_4=7L/10+\Delta
z_{34}$ with $\Delta z_{34}=L/70$. In this example we choose
$L=0.02\,$m.

\h Again, we can calculate the coefficients $A_n$ by inserting
Eq.(\ref{par}) into Eq.(\ref{An}), and use them in our
superposition (\ref{soma2}). In this case, we chose
$Q=0.995\,\om_0/c$: This choice allows a maximum value of $n$ given
by $N_{\rm max}=158$ (one can see this by
exploiting Eq.(\ref{N})). But for simplicity we adopt once more
$N=35$, hoping that Eq.(\ref{soma3}) will yield a good enough
approximation of the desired function.

\h We compare in Fig.2(a) the intensity of the desired
longitudinal function $F(z)$ with that of our FW, \
$\Psi(\rho=0,z,t)$, \ obtained from Eq.(\ref{soma2}) by using
$N=35$: \ We can verify a good agreement between the desired longitudinal
behaviour and our FW. Obviously we can improve the approximation
by using larger values of $N$.

\begin{figure}[!h]
\begin{center}
 \scalebox{1}{\includegraphics{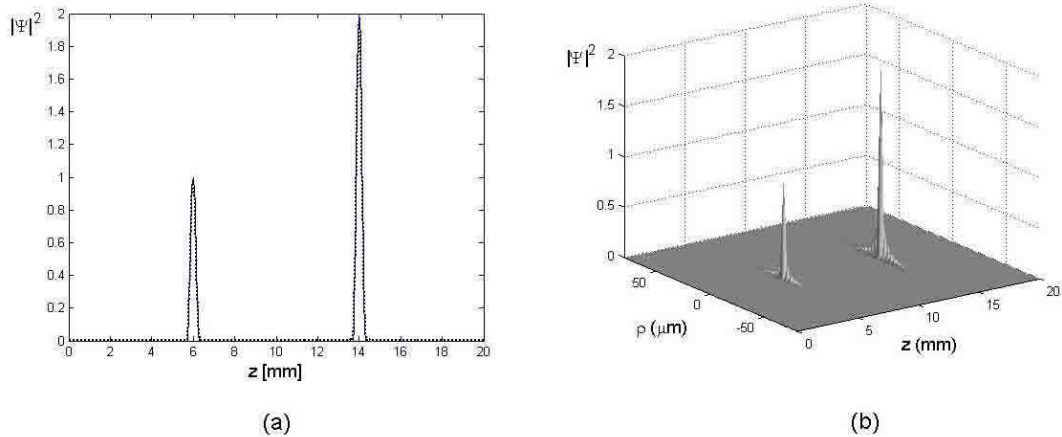}}
\end{center}
\caption{\textbf{(a)}Comparison between the intensity of the
desired longitudinal function $F(z)$, given by Eq.(\ref{par}), and
that of our FW,\ $\Psi(\rho=0,z,t)$, \ obtained from
Eq.(\ref{soma2}). The solid line represents the function $F(z)$,
and the dotted one our FW.\textbf{(b)}3D plot of the field
intensity of the FW chosen by us in this new case.} \label{fig2}.

\end{figure}

\h In Fig.2(b) we show the 3D field intensity of our FW, forwarded
by Eq.(\ref{soma3}). We can see that this field has a good
transverse localization and possesses the desired longitudinal
pattern.

\newpage

\h {\em Third example} (controlling the transverse shape too):

We want to take advantage of this new example for addressing an
important question: \ We can expect that, for a desired longitudinal
pattern of the field intensity, by choosing smaller values of the
parameter $Q$ one will get FWs with narrower {\em transverse}
width [for the same number of terms in the series entering
Eq.(\ref{soma3})], because of the fact that the Bessel beams in
Eq.(\ref{soma3}) will possess larger transverse wave numbers, and,
consequently, higher transverse concentrations. \ We can verify
this expectation by considering, for instance, inside the usual range
$0 \leq z \leq L$, the longitudinal pattern represented by the
function

 \bb
 F(z) \ug \left\{\begin{array}{clr}
 -4\,\,\dis{\frac{(z-l_1)(z-l_2)}{(l_2 - l_1)^2}} \;\;\; &
 {\rm for}\;\;\; l_1 \leq z \leq l_2  \\

 \\
 \;\;\;\;\;\;\;\; 0  & \mbox{elsewhere}
\end{array} \right. \; , \label{Fz2}
 \ee

with $l_1=L/2-\Delta z$ and $l_2=L/2+\Delta z$.  Such a function
has a parabolic shape, with its peak centered at $L/2$ and with
longitudinal width $2 \Delta z/\sqrt{2}$.  By adopting $\lambda_0
= 0.632\;\mu$m (that is, $\om_0 = 2.98 \times 10^{15}\;$Hz), let us
use the superposition (\ref{soma3}) with {\em two} different
values of $Q$: \ We shall obtain two different FWs that, in spite of
having the same longitudinal intensity pattern, will possess
different transverse localizations. Namely, let us consider
$L=0.06\,$m and $\Delta z = L/100$, and the two values
$Q=0.999\,\om_0/c$ and $Q=0.995\,\om_0/c$. In both cases the
coefficients $A_n$ will be the same, calculated from
Eq.(\ref{An}), on using this time the value $N=45$ in
superposition (\ref{soma3}). The results are shown in Figs.3(a)
and 3(b). Both FWs have the same longitudinal intensity pattern,
but the one with the smaller $Q$ is endowed with a narrower
transverse width.

\begin{figure}[!h]
\begin{center}
\scalebox{1}{\includegraphics{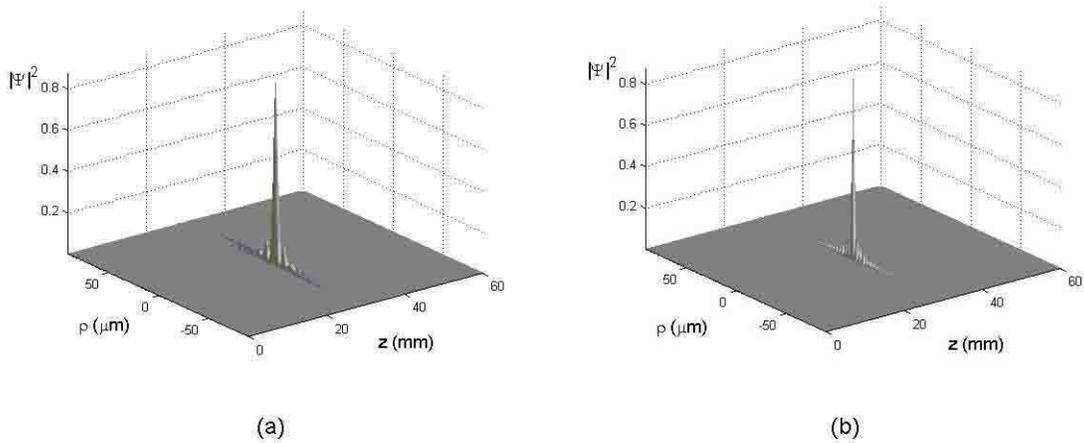}}
\end{center}
\caption{\textbf{(a)} The Frozen Wave with $Q=0.999\om_0/c$ and
$N=45$, approximately reproducing the chosen longitudinal pattern
represented by Eq.(\ref{Fz2}). \ \textbf{(b)} A different Frozen
wave, now with $Q=0.995 \om_0/c$ (but still with $N=45$)
forwarding the same longitudinal pattern. We can observe that in
this case (with a lower value for $Q$) a higher transverse
localization is obtained.} \label{fig5}
\end{figure}

\

\h In Section 5 we shall show that a better control of the
transverse shape can be obtained by using higher order
Bessel beams in superposition (\ref{soma3}).

\

\

\section{Spatial resolution and Residual intensity}

\h In connection with a FW of a given frequency, it is of
practical (and theoretical) interest to investigate its {\em Spatial
resolution}, its {\em Residual intensity}, the {\em Size of the source}
necessary to generate it, as well as the minimum distance from the
source needed to get such a FW.

\h Let us first comment that, in lossless media, the theory of FWs
can furnish results similar to the free-space ones. This happens
because FWs are suitable superpositions of Bessel beams with the
same frequency, so that there is no problem with the material
dispersion.


\h Here, we deal with lossless media only.

\h Let us address the question of the longitudinal {\em and} transverse
spatial resolution for the FWs.

\h In connection with the longitudinal case, once we choose a desired
longitudinal intensity field configuration, $|F(z)|^2$, given, for
example, by a single peak (or a few peaks) with a certain longitudinal width
$\Delta z$, we wish to investigate whether it is possible to obtain such a
spatial resolution, and what are the relevant parameters for getting good
results.

\h As one can expect, this question is directly related to the
number $2N+1$ of terms in superposition (\ref{soma3}): More
specifically, in superposition (\ref{soma2}).

\h Once the values of the frequency $\om_0$, and the parameters
$L$ and $Q$ are chosen, the best approximation that we can get for
a given longitudinal intensity-field configuration, $|F(z)|^2$, is
obtained by using Eq.(\ref{soma2}) with the maximum number of
terms $2N_{\rm max}+1$, where $N_{\rm max}$ is calculated from is
calculated through inequality (\ref{N}).

\h As we have seen in the previous Sections, it is not always
necessary to use $N=N_{\rm max}$, and frequently a smaller value
of $N$ can provide us with good results. But even in this cases, when a value
$N < N_{\rm max}$ is quite sufficient to furnish the desired spatial
resolution, nevertheless it can be desirable to increase the value of $N$ for
lowering the longitudinal residual intensities, as we are going to see.

\h In the cases in which not even the value $N=N_{\rm max}$ yields a
good result, we have to adopt a smaller value for the parameter $Q$
so to increase, in this way, the value of $N_{\rm max}$ itself.
For quantifying mathematically the precision of our approximation, one may
have recourse to the mean square deviation, $D$,

$$ D \ug \int_{0}^{L}\,|F(z)|^2 \,dz  - L\,\sum_{-N}^{N} |A_n|^2 \; ,  $$

where $A_n$, the coefficients of superposition (\ref{soma2}), are
given by Eq.(\ref{An}).

\h The case of the transverse spatial resolution cannot be tackled
in such a detail, since, as we know, one cannot have a complete
three-dimensional control of the field.  In the previous Section
we have seen that we can get, however, some control on the
transverse spot size through the parameter $Q$. Actually,
Eq.(\ref{soma3}), that defines our FW, is a superposition of
zero-order Bessel beams, and, due to this fact, the resulting
field is expected to possess a transverse localization around
$\rho=0$. Each Bessel beam in superposition (\ref{soma3}) is
associated with a central spot with transverse size, or width,
$\Delta\rho_n \approx 2.4/k_{\rho\,n}$. On the basis of the
expected expected convergence of series (\ref{soma3}), we can
estimate the width of the transverse spot of the resulting beam as
being

\bb \Delta \rho \approx \frac{2.4}{k_{\rho\,n=0}} \ug
\frac{2.4}{\sqrt{\om_0^2/c^2 - Q^2}} \; , \label{tspot} \ee

which is the same value as that for the transverse spot of the Bessel beam
with $n=0$ in superposition (\ref{soma3}). Relation (\ref{tspot}) can be
useful: Once we have chosen the desired longitudinal intensity
pattern, \emph{we can choose even the size of the transverse spot,
and use relation (\ref{tspot}) for evaluating the needed, corresponding
value of parameter $Q$.}

\h In spite of the fact that the transverse spot size happens to be
approximately equal to that of a Bessel beam with $k_{\rho} = \sqrt{\om_0^2/c^2 - Q^2}$,
it may happen that the {\em decay} of the field transverse intensity for
$\rho > \Delta\rho$  is much faster than that of an ordinary Bessel
beam!  This happens when the desired field intensity presents a
longitudinal width $\Delta z$ much smaller than $L$, i.e., $\Delta
z << L$, as we will see below.

\

\h {\em An illustrative example:}

Let us consider the situation in which, within the interval $0
\leq z \leq L$, the desired longitudinal intensity pattern is given
by a well-concentrated peak, represented by expression
(\ref{Fz2}), with $\lambda_0 = 0.632\;\mu$m (that is, $\om_0 =
2.98 \times 10^{15}\;$Hz), \ $Q=0.98\,\om_0/c$, \ $L=0.01\,$m, and $\Delta
z = L/500$.

\h Figures 4(a), 4(b) and 4(c) show the resulting FWs obtained by
using, in superposition (\ref{soma2}), $N=100$, $N=250$ and
$N=300$, respectively. We can see in the first case that $N=100$
is not enough for yielding a good result. On the other hand, the
second and third cases, with $N=250$ and $N=300$, seem to
reproduce the desired pattern very well, with no apparent
difference between the two cases. However, Fig.5 shows that the
residual intensity for the third case is {\em smaller} than for
the second one, confirming the previous conclusions.

\h Figure 6(a) shows the transverse intensity pattern of the peak
(in the plane $z=L/2$) for the case with $N=250$. We can see that
the value of the transverse spot width agrees very well with our
estimate (\ref{tspot}), which furnishes, in this case, the value
$\Delta\rho\approx1.22\,\mu$m. \  From Fig.6(b) one can visually
evaluate the residual intensity of the transverse pattern: One can
observe that the transverse decay is {\em strong} and much faster
than that presented by Bessel beams. This figure too confirms our
previous conclusions.

\newpage


\begin{figure}[!h]
\begin{center}
 \scalebox{1}{\includegraphics{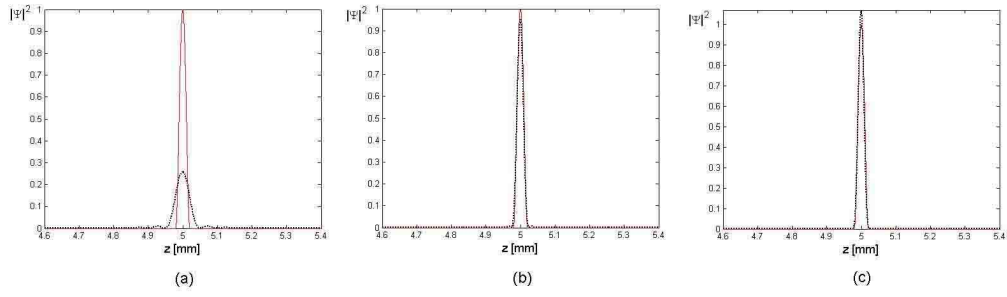}}
\end{center}
\caption{Comparison of the desired longitudinal intensity pattern
(solid line) with those of the resulting FWs (dotted line), when
using: \textbf{(a)} $N=100$; \ \textbf{(b)} $N=250$; \
\textbf{(c)} $N=300$.} \label{Fig4}

\end{figure}

\begin{figure}[!h]
\begin{center}
 \scalebox{1}{\includegraphics{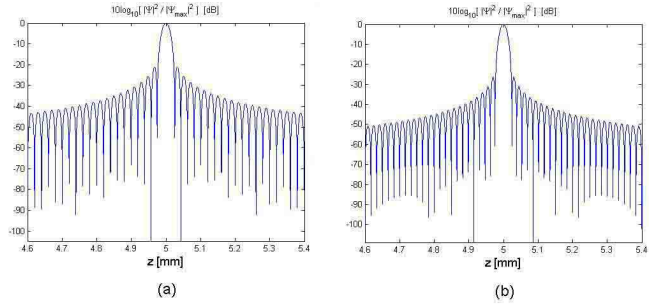}}
\end{center}
\caption{\textbf{(a)} Longitudinal residual intensity of the considered FW
with $N=250$. \ \textbf{(b)} The same with $N=300$.}
\label{Fig5}

\end{figure}

\begin{figure}[!h]
\begin{center}
 \scalebox{1}{\includegraphics{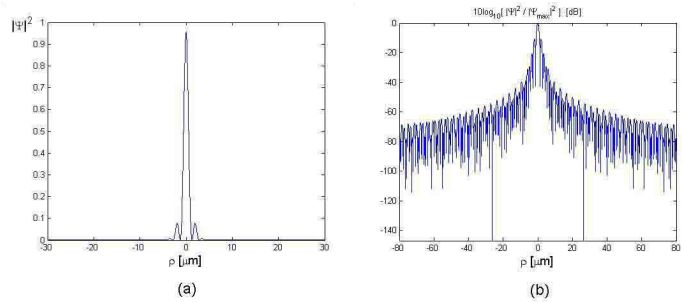}}
\end{center}
\caption{\textbf{(a)} Transverse intensity pattern for the peak of
the considered FW with $N=250$. \ \textbf{(b)} The transverse residual
intensity for this case.} \label{Fig6}

\end{figure}

\newpage

\section{Increasing the control on the transverse shape by using
higher-order Bessel beams}

\h We have shown in the previous Section how to get a very strong
control over the longitudinal intensity pattern of a beam using
suitable superposition of zero-order Bessel beams.

\h As already mentioned, due to the fact that the resulting beam
must obey the wave equation, we cannot get a total
three-dimensional control of the wave pattern; but we have shown
that we {\em can} have some control on the transverse behavior:
More specifically, we can control the transverse spot size through
the parameter $Q$, which defines the values of the {\em transverse
wave numbers} of the Bessel beams entering superposition
(\ref{soma3}).

\h In this Section we are going to argue that it is possible to increase
even more our control of the transverse shape by using
higher-order Bessel beams in our fundamental superposition (\ref{soma3}). \
Despite the method presented {\em in this Section} is not yet
demonstrated in a rigorous mathematical way, it can be understood and
accepted on the basis of simple and intuitive arguments. \ The
basic idea is obtaining the desired longitudinal intensity
pattern, not along the axis $\rho=0$, but on a cylindrical surface
corresponding to $\rho = \rho' > 0$. \ This allows one to get
interesting stationary field distributions, as static annular
structures (tori), or cylindrical surfaces, of stationary light
(or electromagnetic or acoustic field), and so on, with many
possible applications.$^{**}$ \ To realize this, let us initially start with
the same procedure in the previous Section; i.e., let us choose some desired
{\em longitudinal} intensity pattern, within the interval $0 \leq z \leq
L$, and calculate the coefficients $A_n$ by using Eq.(\ref{An}).
Afterwards, let us replace the zero-order Bessel beams $J_0(k_{\rho\,n}\rho)$,
in superposition (\ref{soma3}), with higher-order Bessel beams,
$J_{\mu}(k_{\rho\,n}\rho)$, to get

\bb \dis{\Psi(\rho,z,t) \ug
e^{-i\,\om_0\,t}\,e^{i\,Q\,z}\,\sum_{n=-N}^{N}
A_n\,J_{\mu}(k_{\rho\,n}\,\rho)\,e^{i\,\frac{2\pi}{L}n\,z} } \; ,
\label{soma4} \ee

with $A_n = (1/L) \int_{0}^{L}\,F(z)\,{\rm exp}(-i 2 \pi n z
/L)\,\drm z$, and  $k_{\rho\,n} = \sqrt{\om_0^2 - \left(Q + 2 \pi
n/L \right)^2}$.

\h In superposition (\ref{soma4}), the Bessel functions
$J_{\mu}(k_{\rho\,n}\rho)$, with different values of $n$, reach their
maximum values at $\rho=\rho'_n$, where $\rho'_n$ is
the first positive root of the equation
$(\drm\,J_{\mu}(k_{\rho\,n}\rho)/\drm\rho)|_{\rho'_n}=0$. \ The
values of $\rho'_n$ are located around the central value
$\rho'_{n=0}$, at which the Bessel function
$J_{\mu}(k_{\rho\,n=0}\rho)$ assumes its maximum value. \
We can intuitively expect that the desired longitudinal
intensity pattern, initially constructed for $\rho=0$, will approximately
shift to $\rho = \rho'_{n=0}$. \ We have found such a conjecture
to hold in all situations explicitly considered by us. \ By such a
procedure, one can obtain very interesting stationary
configurations of field intensity, as the mentioned
``donuts" and cylindrical surfaces, and much more.

\h In the following example, we show how to obtain, e.g., a cylindrical
surface of stationary {\em light}. \ To get it, within the interval $0 \leq z
\leq L$, let us first select the longitudinal intensity pattern given by Eq.(\ref{Fz2}),
with $l_1=L/2-\Delta z$ and $l_2=L/2+\Delta z$, and with $\Delta z =
L/300$. Moreover, let us choose $L=0.05\,$m, $Q=0.998\,\om_0/c$,
and use $N=150$.

\h Then, after calculating the coefficients $A_n$ as before,

$$A_n \ug \frac{1}{L} \dis{
\int_{0}^{L}\,F(z)\,e^{-i\,\frac{2\pi}{L}\,n\,z}\,d\,z }\;,$$

we have recourse to superposition (\ref{soma4}). In this case, we
choose $\mu =4$. \ According to the previous discussion, one can
expect the desired longitudinal intensity pattern to appear
shifted to $\rho ' \approx 5.318/k_{\rho\,n=0}=8.47\,\mu$m, where
5.318 is the value of $k_{\rho\,n=0}\,\rho$ for which the Bessel
function $J_{4}(k_{\rho\,n=0}\,\rho)$ assumes its maximum value,
with $k_{\rho\,n=0} = \sqrt{\om_0^2 - Q^2}$. \ The figures below
show the resulting intensity field.

\h Figure 7(a) depicts the transverse intensity pattern for
$z=L/2$. The transverse peak intensity is located at
$\rho=7.75\,\mu$m, with a $8.5\%$ difference w.r.t. the predicted
value of $8.47\,\mu$m. \ In Fig.7(b) the transverse section of the
resulting beam for $z=L/2$ is shown.

\h Figure 8 depicts the three-dimensional pattern of such a higher-order
FW. In Fig.8(a) the orthogonal projection of its 3D pattern is shown,
which corresponds to nothing but a cylindrical surface of stationary light
(or other fields). \ In Fig.8(b) the same field is shown, but from a
different point of view.

\

\begin{figure}[!h]
\begin{center}
 \scalebox{.9}{\includegraphics{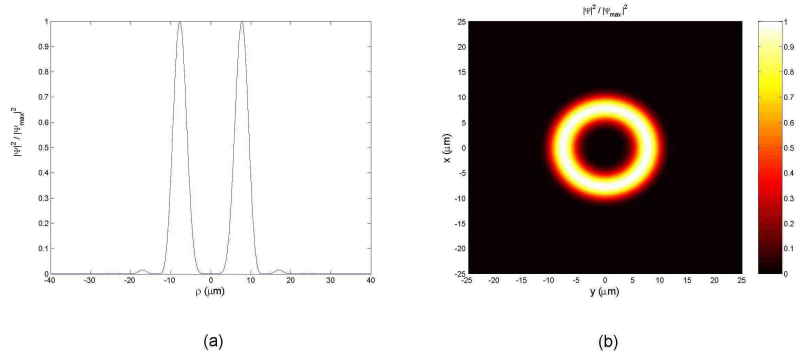}}
\end{center}
\caption{\textbf{(a)} Transverse intensity pattern at $z=L/2$ of
the considered, higher-order FW. \ \textbf{(b)} Transverse section
of the resulting stationary field for $z=L/2$.} \label{Fig7}.

\end{figure}

\

\begin{figure}[!h]
\begin{center}
 \scalebox{.9}{\includegraphics{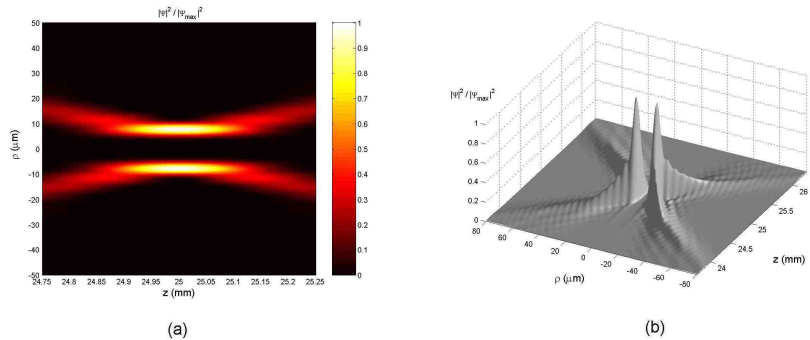}}
\end{center}
\caption{\textbf{(a)} Orthogonal projection of the three-dimensional
intensity pattern of the higher-order FW depicted in Figs.7. \ \textbf{(b)}
The same field but under a different perspective.} \label{Fig8}.

\end{figure}

We can see that the desired longitudinal intensity pattern has
been approximately obtained, but, as wished, shifted from $\rho=0$
to $\rho = 7.75\,\mu$m: and the resulting field resembles a
cylindrical surface of stationary light with radius $7.75\,\mu$m
and length $238\,\mu$m. Donut-like configurations of light (or
sound) are also possible.

\

\

\section{Generation of Frozen Waves}

In the previous Sections, we have shown how suitable
superpositions of Bessel beams of the same frequency can provide
impressive results: Namely, can produce {\em stationary} wavefields
with high transverse localization, and with an arbitrary
{\em longitudinal} shape, within a chosen space interval $0\leq z \leq
L$; that is, Frozen Waves with a static envelope.  As we already mentioned,
such waves are rather interesting, not only from the theoretical
point of view, but also because of their great variety of possible
applications, ranging from ultrasonics to laser surgery, and from tumor
destruction to optical tweezers.

But how to produce our FWs? \ Regarding the generation of
FWs, one has to recall that superpositions (\ref{soma3}), which define
them, consist of sums of Bessel beams. Let us also recall that a Bessel
beam, when generated by a finite aperture (as it must be, in any
real situations), maintains its nondiffracting properties till a
certain distance only (called its field-depth), given by

\bb Z \ug  \frac{R}{\tan\theta} \; , \label{ldif}\ee

where $R$ is the aperture radius and $\theta$ is the so-called
axicon angle, related to the longitudinal wave number by the
known expression $\cos\theta=c\beta/\om$.

\h So, given an apparatus whatsoever capable of generating a single
(truncated) Bessel beam, we can use {\em an array} of such
apparata to generate a sum of them, with the appropriate
longitudinal wave numbers and amplitudes/phases [as required by
Eq.(\ref{soma3})], thus producing the desired FW. \ Here, it is
worthwhile to notice preliminarily that we shall be able to generate
the desired
FW in the the range $0\leq z \leq L$ if all Bessel beams entering
the superposition (\ref{soma3}) are able to reach the distance
$L$ resisting the diffraction effects. We can guarantee this,
for instance, if $L
\leq Z_{\rm min}$, where $Z_{\rm min}$ is the field-depth of the
Bessel beam with the smallest longitudinal wave number
$\be_{n=-N}=Q-2\pi N/L$, that is, with the shortest depth of
field. In such a way, once we have the values of $L$, $\om_0$,
$Q$, $N$, from Eq.(\ref{ldif}) and from the above considerations it
results that the radius $R$ of the finite aperture has to be

\bb R \geq L\dis{\sqrt{\frac{\om_0^2}{c^2\be_{n=-N}^2}-1}} \ee

\h The simplest apparatus capable of generating a Bessel beam is
that adopted by Durnin et al.\cite{du}, which consists in an
annular slit located at the focus of a convergent lens and
illuminated by a cw laser. Then, an array of such annular rings
with the appropriate radii and transfer functions, able to yield
both the correct longitudinal wave numbers\footnote{Once a value
for $Q$ has been chosen.} and the coefficients $A_n$ of the
fundamental superposition (\ref{soma3}), can generate the desired
FW.

\ In the next Section we shall just consider such a simple
apparatus, even if, of course, other powerful tools, like the
computer generated holograms, may be used to produce the FWs.

\

\subsection{A very simple apparatus for producing FWs}

Let us work out an example, by having recourse to {\em an array}
of the very simple Durnin et al.'s experimental apparata.

\h Since 1987, let us repeat, it has been used a simple
experimental mean for creating a Bessel beam, consisting in an
annular slit located at the focus of a convergent lens and
illuminated by a cw laser. Let us call $\delta a$ the width of the
annular slit, $\la$ the wavelength of the laser, and $f$ and $R$
the focal length and the aperture radius of the lens,
respectively. On illuminating the annular slit with a cw laser
with frequency $\om_0$, and provided that condition $\delta a \ll
\la f/R$ is satisfied, the Durnin et al.'s apparatus creates,
after the lens, a wavefield closely similar to a Bessel beam
along a certain depth of field.  Within such field-depth,
$z<R/\tan\theta$, and to $\rho<<R$, the generated Bessel beam can
be approximately written

\

\bb \psi(\rho,z,t)\approx \La J_0(k_{\rho}\rho)\,e^{i\be
z}e^{i\om_0 t} \label{bbd} \ee

\

with $\La$ a constant depending on the values of $a$, $f$, $\om_0$
and $\delta a$,

\

\bb k_{\rho} \ug \frac{\om_0}{c}\frac{a}{f} \; , \ee

and

\bb \be^2 \ug \frac{\om_0^2}{c^2} - k_{\rho}^2 \; . \ee

\

Thus, as Durnin et al. suggested, we can see that the transverse and
longitudinal wave numbers are determined by radius and focus
of slit and lens, respectively. \ Once more, let us recall also that the wavefield has approximately a
Bessel beam behavior (when $\rho<<R$), in the range $0 \leq z \leq
Z \approx R\,f/a $ that we have called the field-depth of the Bessel
beam. \ Our FWs are to be obtained by suitable superpositions of Bessel
beams. So we can experimentally produce the FWs by using several
concentric annular slits (Fig.9), where each radius is chosen in order
yield the correct longitudinal wave number, and where the transfer
function of each annular slit is chosen in order to furnish the
coefficients $A_n$ of Eq.(\ref{soma2}) which are needed for
the desired longitudinal pattern to be obtained.

\

\begin{figure}[!h]
\begin{center}
 \scalebox{1.1}{\includegraphics{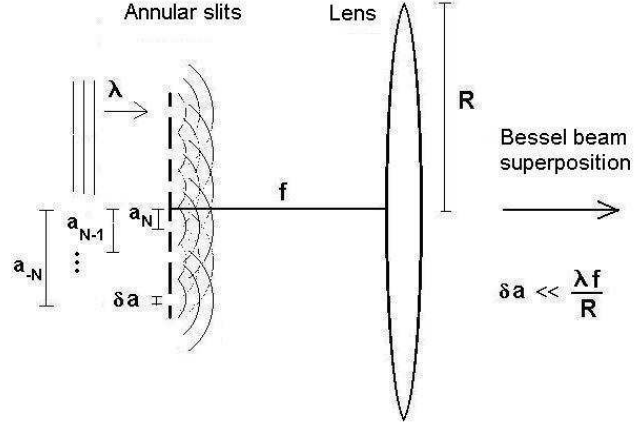}}
\end{center}
\caption{A set of suitable, concentric annular slits, as a simple means
for generating a Frozen Wave.}
\label{fig8}
\end{figure}

\h Let us examine all this in more details. \ Suppose we have
$2N+1$ concentric annular slits with their radii given by $a_n$, with \
$-N \leq n \leq N$.  \ Along a certain range, after the lens, one
will have a wavefield given by the sum of the Bessel beams
produced by each slit, namely\footnote{The same apparatus could
also be used to generate higher order FWs, when the zero-order
Bessel beams in superposition (\ref{soma5}) are replaced with
higher order Bessel functions. Experimentally, it can be performed by
angular modulation of the slits.}

\

\bb \dis{\Psi(\rho,z,t) \ug e^{-i\,\om_0\,t}\,\sum_{n=-N}^{N}
\La_n\, T_n \,J_0(k_{\rho\,n}\rho)\,e^{i\,\be_n\,z} } \; ,
\label{soma5} \ee

\

$T_n$ being the transfer function of the $n$-th annular slit
(which regulates amplitude and phase of the emitted Bessel beam,
and is a constant function for each slit); while the $\La_n$ are
constants depending on the characteristics of the apparatus:
Namely depending, in general, on the values of $a$, $f$, $\om_0$
and $\delta a$. It is possible to obtain a simple expression for
the $\Lambda_n$ by making some simplifying, rough
considerations\cite{cesar}. The transverse and longitudinal wave
numbers are given by

\bb k_{\rho n} \ug \frac{\om_0}{c}\frac{a_n}{f} \label{kn} \ee

and

\bb \be_n^2 \ug \frac{\om_0^2}{c^2} - k_{\rho}^2 \; . \label{bn} \ee

\

\h On the other hand, we know from the present theory that for constructing
the FWs quantity $\be$ is to be given by Eq.(\ref{be2}):

\

$$ \be_n \ug Q + \frac{2\,\pi}{L}\,n \; . $$

\

\h On combining Eqs.(\ref{be2},(\ref{kn}),(\ref{bn}), one gets

\

\bb \left(Q + \frac{2\,\pi}{L}\,n \right)^2 \ug
\frac{\om_0^2}{c^2} - \left(\frac{\om_0}{c}\frac{a_n}{f}\right)^2
\ee

\

and, solving with respect to $a_n$,

\

\bb a_n \ug f \dis{\sqrt{1-\frac{c^2}{\om_0^2}\left(Q +
\frac{2\,\pi}{L}\,n \right)^2}} \; . \label{an} \ee

\

\h Equation (\ref{an}) yields the radii of all the annular slits
that provide the correct longitudinal wave numbers, needed for the
generation of the FWs. We may notice that the radii of the annular
slits do not depend on the specific desired longitudinal
intensity-pattern, and that many different sets of values for the radii
are possible on making different choices for the parameter $Q$.

\h Notice that the procedure is not yet finished.  Indeed, once
the desired longitudinal pattern $F(z)$ has been chosen, one has
necessarily to meet in Eq.(\ref{soma2}) the coefficients $A_n$
given by Eq.(\ref{An}); and such coefficients have to be the
coefficients of Eq.(\ref{soma5}). For obtaining them, it is
necessary that each annular slit be endowed by the appropriate
transfer function, which regulates {\em amplitude and phase} of
the Bessel beam emitted by that slit. \ By using
Eqs.(\ref{An},(\ref{soma3}),(\ref{soma5}), we get the transfer
function $T_n$ of the $n$-th annular slit to be

\

\bb T_n \ug \frac{A_n}{\La_n} \ug \frac{1}{L\,\La_n}\,\dis{
\int_{0}^{L}\,F(z)\,e^{-i\,\frac{2\pi}{L}\,n\,z}\,\drm\,z } \; ,
\label{Tn}\ee

\

\h Finally, with the radius of each annular slit given by
Eqs.(\ref{an}) and the transfer functions of each slit given by
Eqs.(\ref{Tn}), we do obtain a FW endowed with the desired
longitudinal behaviour, inside the interval $0 \leq z \leq L$.
Of course, one has to guarantee also that the distance $L$ is smaller
than the smallest field-depth of the Bessel beams entering
superposition (\ref{soma5}). In other words, one must have also

\

\bb L \leq Z_{\minrm} \approx \frac{R\,f}{a_{\maxrm}} \ee

\

where $a_\maxrm$ is the largest radius of the concentric annular
slits.

\

\

\section{Conclusions}

\h In this work we have expounded the theory of Frozen Waves, and depicted
some possible experimental apparata to generate them.
The present results can find applications in many fields: Just to make
an example, in optical tweezers modelling, since we can
construct stationary optical (but also acoustic, etc.) fields with a great
variety of shapes; capable, e.g., of trapping particles or tiny objects at
different locations.$^{**}$ \ These topics will be reported elsewhere.

\

\

\section*{Acknowledgements}

The authors are very grateful, for collaboration and many
stimulating discussions over the last few years, with Marco
Mattiuzzi. This work has been partially supported by FAPESP
(Brazil), and by INFN, MIUR and Bracco Imaging SpA (Italy). Thanks
are also due for stimulating discussions to M.Brambilla, C.Cocca,
and G.Degli Antoni.

\end{document}